\documentclass[a4paper,fleqn,usenatbib]{mnras}
\usepackage{color}
\usepackage[T1]{fontenc}
\usepackage{aecompl}
\usepackage{natbib}

\bibpunct{(}{)}{;}{a}{,}{,}
\usepackage[british]{babel}
\usepackage[T1]{fontenc}
\usepackage{ae,aecompl}
\newcommand{\degree}{\ensuremath{^\circ}}

\usepackage{graphicx}	
\usepackage{amsmath}	
\usepackage{amssymb}	
\usepackage{txfonts,comment}

\title[Spectral and Time Series Analyses of Zw 229.015]{Spectral and Time Series Analyses of the Seyfert 1 AGN: Zw 229.015}

\author[Adegoke, Rakshit \& Mukhopadhyay]
{Oluwashina Adegoke$^1$\thanks{E-mail : adegoke@physics.iisc.ernet.in}, Suvendu Rakshit$^2$
\thanks{E-mail : suvendu.rakshit@iiap.res.in}, Banibrata Mukhopadhyay$^1$\thanks{E-mail : bm@physics.iisc.ernet.in}\\
$^1$Astronomy and Astrophysics Programme, Department of Physics, Indian Institute of Science, Bangalore 560012, India\\
$^2$Indian Institute of Astrophysics, Bangalore 560034, India\\
}
\begin{document}
\maketitle
\label{firstpage}
\begin{abstract}
We analyse the spectra of the archival \textit{XMM-Newton} data of the Seyfert  1 AGN Zw 229.015 in the energy range $0.3-10.0 \,\mathrm{keV}$. When fitted with a simple power-law, the spectrum shows signatures of weak soft excess below $1.0 \,\mathrm{keV}$. 
We find that both thermal comptonisation and relativistically blurred reflection models provide the most acceptable spectral fits with plausible physical explanations to the origin of the soft excess than do multicolour disc blackbody and smeared wind absorption models. This motivated us to study the variability properties of the soft and the hard X-ray emissions from the source and the relationship between them to put further constraints on the above models. Our analysis reveals that the variation in the $3.0-10.0\,\mathrm{keV}$ band lags that in the $0.3-1.0\,\mathrm{keV}$ by ${600^{+290}_{-280}}\,\mathrm{s}$, while the lag between the $1.0-10.0\,\mathrm{keV}$ and $0.3-1.0\,\mathrm{keV}$ is ${980^{+500}_{-500}}\,\mathrm{s}$. This implies that the X-ray emissions are possibly emanating from different regions within the system. From these values, we estimate the X-ray emission region to be within $20R_{\mathrm{g}}$ of the central supermassive black hole (where $R_{\mathrm{g}}={GM}/c^{2}$, $M$ is the mass of black hole, $G$ the Newton's gravitational constant and 
$c$ the speed of light). Furthermore, we use \textit{XMM-Newton} and \textit{Kepler} photometric lightcurves of the source to search for possible nonlinear signature in the flux variability. We find evidence that the variability in the system may be dominated by stochasticity rather than deterministic chaos which has implications for the dynamics of the accretion system.


\end{abstract}
\begin{keywords} 
galaxies: active - galaxies -  Seyfert - X-ray - individual:Zw 229.015 
\end{keywords}

\section{Introduction}
The luminosity of an active galactic nucleus (AGN) originates 
from an accretion disc around the supermassive black hole usually in 
the optical/UV/EUV energy bands. A hard power-law generally dominates 
above $2.0\, \mathrm{keV}$ and is believed to emanate from a hot corona or a sub-Keplerian flow \citep{2010ApJ...713..105B} made up 
of very high energy plasma $(\lesssim150 \,\mathrm{keV})$ whose geometry is still not well understood. This means the optical/UV photons from the disc undergo inverse Compton scattering into X-ray energies and as such, part of this energy in turn illuminates the disc and the other part is seen as the direct power-law. The part that illuminates the disc is either subsequently reflected or thermalised \citep{1993ARA&A..31..717M}. AGNs have been known to vary in their luminosities in all wavelengths and the study of the underlying mechanisms have proved to be efficient in helping to understand the central region of AGNs.

Zw 229.015 is a Seyfert 1 AGN at a redshift $z=0.028$ having a low galactic absorption column density ($N_{H}\approx6.25\times10^{20}\mathrm{cm}^{-2}$). Despite being relatively bright, its position in the galactic plane has resulted in there 
being very little known about this source \citep{2012ApJ...749...70C}. Its Seyfert nature was noted by \citet{1990IAUC.5134....2P} and it was observed in the ROSAT all sky survey by \citet{2001A&A...378...30Z}. \citet{2011ApJ...732..121B} carried out a ground based reverberation mapping campaign on the source to complement its scheduled observation with the \textit{Kepler} mission. They detected strong variability in the source exhibiting more than a factor of 2 change in its  broad H$\beta$ flux with H$\beta$ full width at half maximum (FWHM) of $2260\pm65\, \mathrm{km~s^{-1}}$. By combining 
the measured H$\beta$ lag with the broad H$\beta$ width measured from the \textit{rms} variability spectrum, they obtained a virial estimate of the mass of the supermassive black hole to be $\sim 10^7M_{\odot}$. More so, they obtained an estimate of its bolometric luminosity $L_{\mathrm{bol}}\sim6.4\times10^{43}\mathrm{erg \,s^{-1}}$ and $L_{\mathrm{bol}}/L_{\mathrm{Edd}}\approx0.05$.  \citet{2011ATel.3484....1E} reported a return of the source to its prior X-ray flux by July 2011 after an earlier report of a dramatic decrease in its flux variability in early June 2011 \citep{2011ATel.3458....1M}. These observations reveal that the source is highly variable. 

\citet{2014ApJ...795....2E} utilized two methods of power spectral analysis to investigate its optical variability and search for the evidence of a bend frequency associated with a characteristic optical variability timescale. They found it to be $\sim5$ days using the full \textit{Kepler} data set of the source. It should be noted that Zw 229.015 exhibits strong and rapid variability in comparison with 
many other objects having a similar black hole mass and luminosity. It is also one of the few 
low-redshift Seyfert galaxies being monitored by \textit{Kepler} with over three years of 
continuous observation. Hence, it is likely that Zw 229.015 will be a very important source 
for studying AGN properties. \textit{Kepler} has been observing $\sim115 \,\mathrm{deg^2}$ of the sky, monitoring $\sim165,000$ sources every 29.4 minutes with unprecedented stability ($\leq0.1\%$ error for a 15$^{\mathrm{th}}$ magnitude source) and high ($>90\%$) duty cycle over a period of several years \citep{2011ATel.3458....1M}. With respect to timing and time series analysis, \citet{2015A&A...576A..17B} used the correlation integral (CI) method to search for signatures of chaos in the AGN W2R 1926+42 using its \textit{Kepler} lightcurves. They reported the system to be stochastic. 
However, this time series method has not been applied to Zw 229.015. 

In this paper, we report the result of the spectral analysis of Zw 229.015 using its 
\textit{XMM-Newton} X-ray data which, to the best of our knowledge, has not been properly reported in the literatures prior to this time. We also report the result of our timing analyses of the source using its \textit{XMM-Newton} and \textit{Kepler} lightcurves. We assume a cosmology with $H_{0} = 70 \,\mathrm{km \, s^{-1} \, Mpc^{-1}}, \Omega_{M} = 0.3, \Omega_{\Lambda} = 0.7$. Errors are quoted at the 90\% confidence level for one parameter of interest unless otherwise mentioned. 

This paper is structured as follows. In section \ref{sec:sec2}, we describe the observation and data reduction procedure. Section \ref{sec:sec3} focuses on the spectral analysis, section \ref{sec:sec4} reports the time series analyses from both the \textit{XMM-Newton} X-ray data and \textit{Kepler}'s optical data and section \ref{sec:sec5} centres on discussions based on our analyses. Finally, in section \ref{sec:sec6} we present a summary of our work.

\section{Observations and Data Reduction}\label{sec:sec2}
Zw 229.015 was observed on June 5, 2011 by \textit{XMM-Newton} operated with medium filter. The observations with the EPIC-PN \citep{2001A&A...365L..27T} and MOS \citep{2001A&A...365L..18S} detectors were of $\sim27 \,\mathrm{ks}$ and $\sim29 \,\mathrm{ks}$ durations respectively in all large window mode. We use the EPIC data for our analysis because it covers the energy range of interest to us ($0.15-12.0 \,\mathrm{keV}$). Table \ref{table:Details of Observation} shows the details of the \textit{XMM-Newton} observation. Although the addition of the MOS data to the spectral fit does not significantly improve the precision with which spectral parameters are determined, we decide to include it for completeness.

\begin{table}
\caption{Details of Observation} 
\centering 
\begin{tabular}{l l} 
\hline\hline 
\textbf{\textit{XMM-Newton}} & \textbf{Parameters} \\ [0.5ex] 
\hline 
Observation ID & 0672530301 \\  
Start Time for PN & 2011-06-05 14:07:22 \\ 
Stop Time for PN & 2011-06-05 21:35:10 \\ 
Start Time for MOS1 & 2011-06-05 13:39:02 \\ 
Stop Time for MOS1 & 2011-06-05 21:39:16 \\ 
Start Time for MOS2 & 2011-06-05 13:39:04\\ 
Stop Time for MOS2 & 2011-06-05 21:39:16 \\ 
EPIC filter & Medium \\ 
Mode(PN) & PrimeLargeWindow \\
Mode(MOS) & PartialPrimeW3\\  [1ex] 
\hline 
\end{tabular}
\label{table:Details of Observation} 
\end{table}

Data reduction for the \textit{XMM-Newton} data follows standard procedure using the Science Analysis System software (SAS version 14.0.0) with updated Current Calibration Files (CCFs). Event files for the PN and MOS detectors are generated using the tasks EPPROC and EMPROC respectively. These tasks allow for calibration of the energy and the astrometry 
of the events registered in each CCD chip and combines them into a single data file. Event file list is extracted using the SAS task EVSELECT. Data from the PN and MOS cameras are screened individually for time intervals with high background when the total count rate in the instrument exceeds 0.35 and 0.40 counts/s for the MOS and 
PN detectors respectively. Good time interval (gti) files are generated after the intervals of the identified flaring particle background have been excluded based on the mentioned count rate cut off criteria. Cleaned event lists are subsequently obtained in line with the SAS standard procedure. Source photons are extracted from a circular region of radius 40$\arcsec$ around the position of the source. Background is estimated from an annulus surrounding the source in each CCD. The X-ray spectra and lightcurves are generated using EVSELECT and background scaling factors are obtained using the task BACKSCALE. The ancillary and photon redistribution matrices are computed using the SAS task ARFGEN and RMFGEN respectively. The resulting spectra are then grouped to have a minimum of 20 counts per bin to facilitate use of the $\chi^{2}$ minimisation technique. Finally, the MOS1 and MOS2 spectra are added to generate a single spectrum with the ADDASCASPEC utility.

The \textit{Kepler} data used in timing analysis are obtained during quarters $3-17$ corresponding to the period between March, 2010 and May, 2013. \textit{Kepler's} long cadence and short cadence lightcurves have integration time of $\sim30$ minutes and $\sim1$ minute respectively. The \textit{Kepler} pipeline \citep{2010ApJ...713L..87J} operates on original spacecraft data to produce calibrated pixel data \citep{2010SPIE.7740E..1XQ}. The next step involves the extraction of the SAP-flux and finally the PDC-flux in which the lightcurves have been conditioned for transit searches. Therefore we use the Simple Aperture Photometric (SAP) data in which long term trends have not been removed from the lightcurves.

\begin{figure}
\includegraphics[width=0.36\textwidth, height=0.35\textheight, angle=-90]{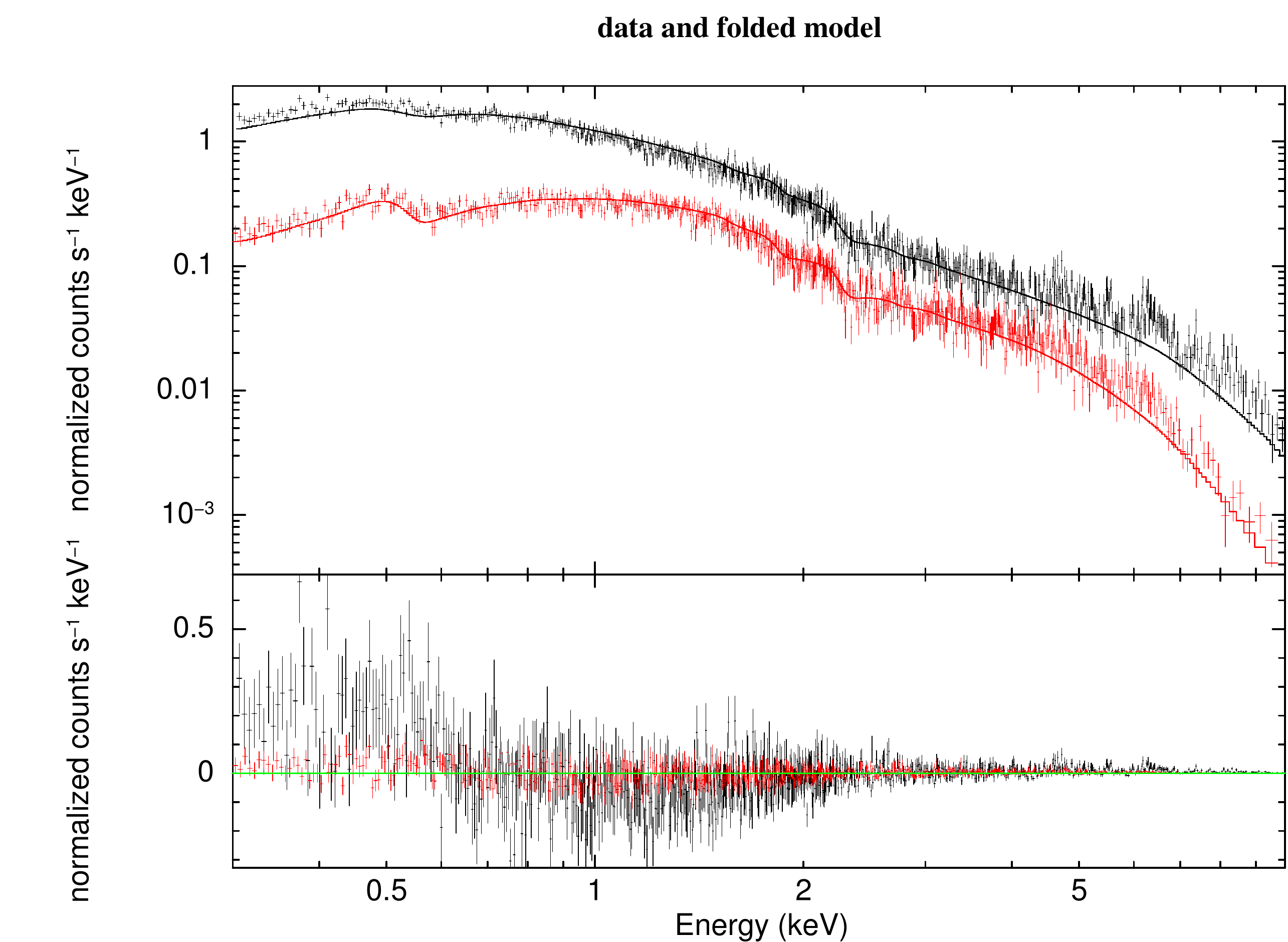}
\caption{The combined EPIC-PN (black upper) and MOS (red lower) spectra of Zw 229.015 fitted with a power-law. Signatures of some soft excess is seen below $1.0 \,\mathrm{keV}$ as evident in the residue (lower) plot.}
\label{Fig:Fig1}
\end{figure}

\section{Spectral Analysis}\label{sec:sec3}
Using XSPEC\footnote{\url{https://heasarc.gsfc.nasa.gov/xanadu/xspec/}} \citep{1996ASPC..101...17A}, we fit the \textit{XMM-Newton} EPIC-PN and MOS spectra with a power-law distribution. We use the absorption model WABS \citep{1983ApJ...270..119M} to account for the galactic absorption of X-rays. We set the equivalent hydrogen column density $N_{H}$ to $6.25 \times 10^{20} \,\mathrm{cm}^{-2}$ as obtained from the column density calculator \citep{2005A&A...440..775K}. We keep this value fixed for all spectral analysis presented in this paper. During the fitting process, we introduce a normalisation constant to account for the relative cross-calibration of the detectors. We carry out the fitting in the energy range $0.3-10.0 \,\mathrm{keV}$. Although the power-law fits the spectra accurately well in the energy range $1.0-10.0 \,\mathrm{keV}$, when extrapolated to lower energies, we notice the presence of a weak soft excess, not uncommon though, it is especially weak for this source. We find the  photon index to be $2.04\pm0.06$ and the $\chi^2/dof$ value to be 2050/1410. The spectra is shown in Figure \ref{Fig:Fig1}. Because the origin of the soft excess is still ill-understood, we set out to use four different physical models to fit the spectra and see which gives the best possible explanation for the origin of the soft excess. The results of fitting with the different models are given below.

\subsection{Multicolour Disc Blackbody Model}\
We fit the soft excess with a multicolour disc blackbody and a power-law using the DISKBB model \citep{1984PASJ...36..741M,1986ApJ...308..635M}. Here we consider the soft excess to result from an optically thick, geometrically thin Shakura-Sunyaev accretion disc \citep{1973A&A....24..337S} around the AGN. The power-law photon index and the temperature of the inner disc are kept as free parameters. On fitting, the photon index obtained is $1.80\pm0.01$ implying a hard power-law and the temperature of the inner disc is $156\pm4 \,\mathrm{eV}$. This is however in agreement with literatures \citep[e.g.,][]{2004MNRAS.349L...7G,2014MNRAS.440..106B}. The $\chi^2/dof$ value obtained is 1491/1408 which implies a reasonably good fit, but the inner disc temperature appears to be quite high and may be absurd
for a Shakura-Sunyaev disc. This is represented in Figure \ref{fig2} (upper panel, left). Table \ref{table:Best-fit parameters} shows the detailed model parameters. 

\begin{table*}
\caption{The best-fit parameters and values from the spectra fitting of the \textit{XMM-Newton} data of Zw 229.015} 
\centering 
\begin{tabular}{l l} 
\hline\hline 
\textbf{Model/Parameter} \& \textbf{Best-fit values} \\ [0.5ex] 
\hline\hline 
Model & constant*wabs*(diskbb+zpo)\\
Calibration factor (CF) & $0.993\pm0.008$\\
$T_{\mathrm{in}}(\mathrm{keV})$ & $0.156\pm0.004$ \\  
Photon index $\Gamma$ & $1.80\pm0.013$ \\ 
$\chi^2/dof$ & 1491/1408 \\ 
Flux ($\mathrm{erg\,cm^{-2}\,s^{-1}}$) & $6.66\times10^{-12}$\\
$L$ ($\mathrm{erg\,s^{-1}}$) & $1.14\times10^{43}$\\
\hline
Model & constant*wabs*(swind1*zpo)\\
Calibration factor (CF) & $0.993\pm0.008$\\
Col. density ($\times10^{22}\,\mathrm{cm}^{-2}$) & $13.80\pm2.27$ \\  
Log($\xi/\mathrm{erg\,cm\,s^{-1}}$) & $3.11\pm0.07$\\
 $\sigma$(in units of $v/c$) & $0.38\pm0.04$\\
Photon index $\Gamma$ & $1.96\pm0.01$ \\ 
$\chi^2/dof$ & 1457/1407 \\ 
Flux ($\mathrm{erg\,cm^{-2}\,s^{-1}}$) & $6.72\times10^{-12}$\\
$L$ ($\mathrm{erg\,s^{-1}}$) & $1.15\times10^{43}$\\
\hline 
Model & constant*wabs*kdblur(atable\{reflionx.mod\}+zpo)\\
Calibration factor (CF) & $0.991\pm0.008$\\
kdblur index & $4.22\pm1.45$ \\  
$R_{\mathrm{in}}(\frac{GM}{c^2})$ & $3.999\pm0.491$\\
Inclination(deg) & 30\\
Photon index $\Gamma$ & $1.52\pm0.082$ \\ 
Fe/Solar & $0.469\pm0.095$\\
Reflionx Xi $(\mathrm{erg\,cm\,s^{-1}})$ & $2308.32\pm658.28$\\
$\chi^2/dof$ & 1436/1404 \\ 
Flux ($\mathrm{erg\,cm^{-2}\,s^{-1}}$) & $6.68\times10^{-12}$\\
$L$ ($\mathrm{erg\,s^{-1}}$) & $1.16\times10^{43}$\\
\hline 
Model & constant*wabs*(comptt+zpo) \textit{disc approximation}\\
Calibration factor (CF) & $0.993\pm0.008$\\
$T_{0}(\mathrm{keV})$ & 0.03 \\  
$KT_{e}(\mathrm{keV})$ & $0.632\pm0.232$\\
Optical depth $\tau_{p}$ & $8.72\pm1.63$\\
Photon index $\Gamma$ & $1.52\pm0.10$ \\ 
$\chi^2/dof$ & 1435/1407 \\ 
Flux ($\mathrm{erg\,cm^{-2}\,s^{-1}}$) & $6.83\times10^{-12}$\\
$L$ ($\mathrm{erg\,s^{-1}}$) & $1.17\times10^{43}$\\
\hline 

\end{tabular}
\label{table:Best-fit parameters} 
\end{table*}

\subsection{Smeared Wind Absorption Model}
This model considers an atomic origin for the soft excess emission. As explained by \citet{2004MNRAS.349L...7G}, the soft excess could be a consequence of smeared absorption from partially ionized material \citep{2007ApJ...671.1284D}. The explanation being that the soft excess is the result of smeared absorption from a partially ionized wind in an accretion disc. To assess this possibility, we fit the data with the smeared wind absorption model SWIND1 in XSPEC. The parameters of this model include absorption column density, ionization parameter $\xi=L/nr^{2}$, ($L$ is the luminosity of the radiation, $r$ is the radius of the disc and $n$ is the number density of the absorbing wind) and the Gaussian velocity dispersion of the wind ($\sigma$) in units of $v/c$. The full list of model parameters and their values are given in Table \ref{table:Best-fit parameters} and Figure \ref{fig2} (upper panel, right) shows the fitted spectrum with the residue. This model provides a good fit to the data with $\chi^{2}/dof=1457/1407$. However, the smearing velocity ($\sigma \sim 0.4c$) required to smoothen the absorption lines to produce the observed soft excess is unphysically large. Similar results have been obtained when the model is applied to some other sources \citep[e.g.,][]{2007ApJ...671.1284D,2014MNRAS.440..106B}. Through self-consistent simulations, \citet{2007MNRAS.381.1413S,2008MNRAS.386L...1S} and \citet{2009ApJ...694....1S} deduced that it is difficult to generate such winds (with sufficient blurring to give the observed soft excess) from an accretion disc driven by radiation or thermal pressure. On the other hand, magnetohydrodynamic simulations show that matter dominated centrifugally driven jets with large opening angles can be self-consistently produced from the accretion disc \citep{2006ApJ...641..103H}. However, as inferred by \citet{2014MNRAS.440..106B}, due to the low densities of such jets ($\sim500\,\mathrm{cm^{-3}}$) sufficient absorption of radiation may not be possible.

\subsection{Relativistically Blurred Reflection Model}
This model posits that the soft excess component results from X-ray ionized reflection. In the model, back-scattering and fluorescence of X-rays in the disc coupled with radiative recombination, causes elements with low ionization potentials, such as C, O, N, to become highly ionized. The fluorescent lines are broadened and blurred due to the high velocities of matter and strong gravitational effects in the inner region of the disc, thus the soft excess is a series of similarly broadened emission lines that blend together to give a continuous emission feature. The model was developed by \citet{2005MNRAS.358..211R}. In XSPEC, the component is implemented as a tabular model REFLIONX and convolved with a LAOR \citep{1991ApJ...376...90L} to account for the blurring of the emission lines from the disc around a spinning black hole, since massive black holes in galactic nuclei are likely to be rapidly spinning \citep{2005ApJ...620...69V,2006MNRAS.365.1067C}. This model assumes a lamp post geometry in which the compact corona located above the black hole illuminates the accretion disc resulting in radial dependent irradiation. Thus the emissivity associated with the reflection is parameterised as $\varepsilon(r) \propto r^{-q}$ where $q$ is the emissivity index. Far away from the black hole, the irradiation decreases as $r^{-3}$, but the light bending effect is strong in the central region focusing some fraction of the coronal emission. Over the radial extent of the disc, the emissivity law can be approximated as a broken power-law \citep{2012MNRAS.419..116F}. 

\begin{figure*}
\centering
\begin{tabular}{@{}cc@{}}
\includegraphics[width=.36\textwidth, angle=-90]{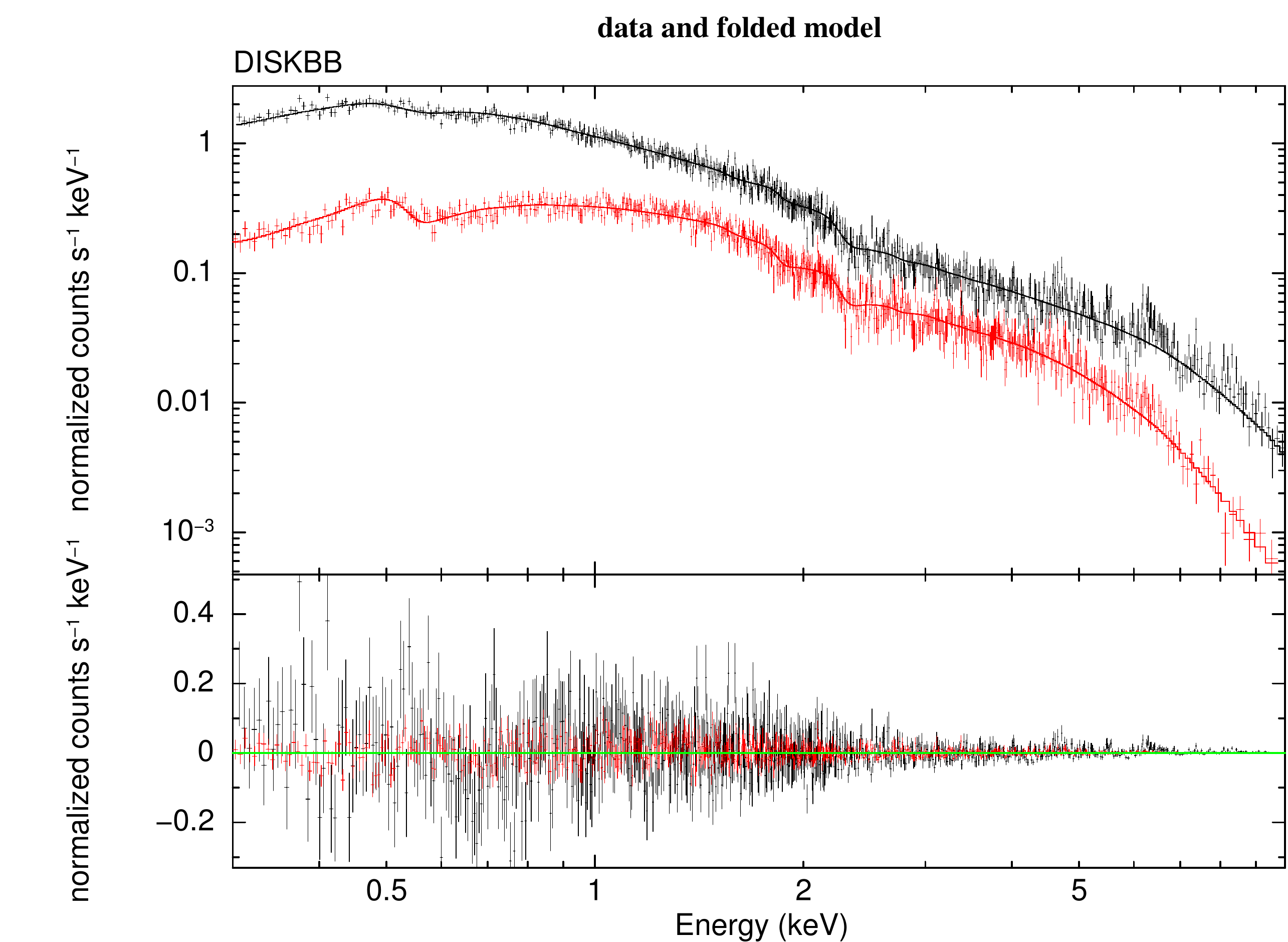} &
\includegraphics[width=.36\textwidth, angle=-90]{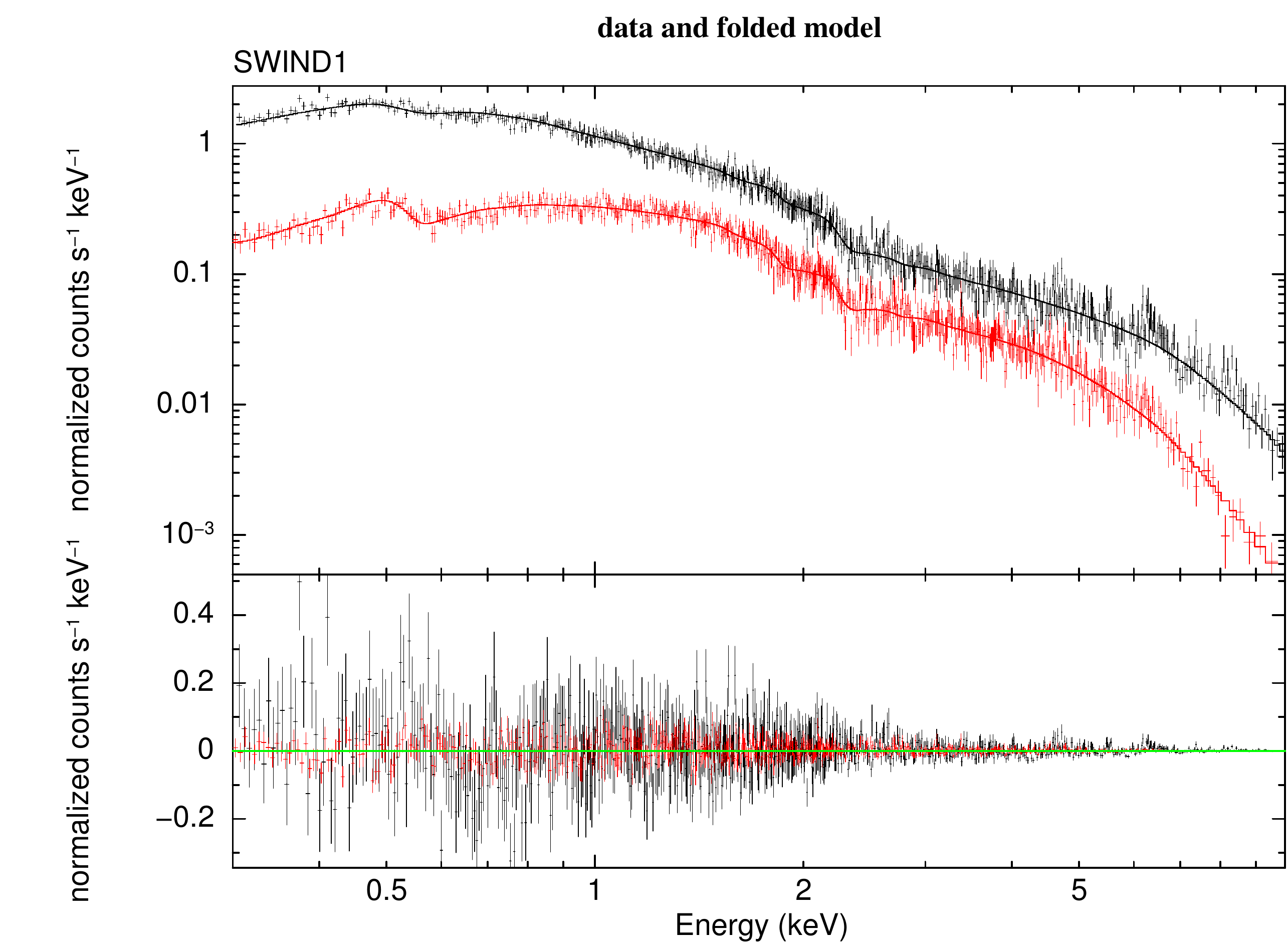} \\
\\
\includegraphics[width=.36\textwidth, angle=-90]{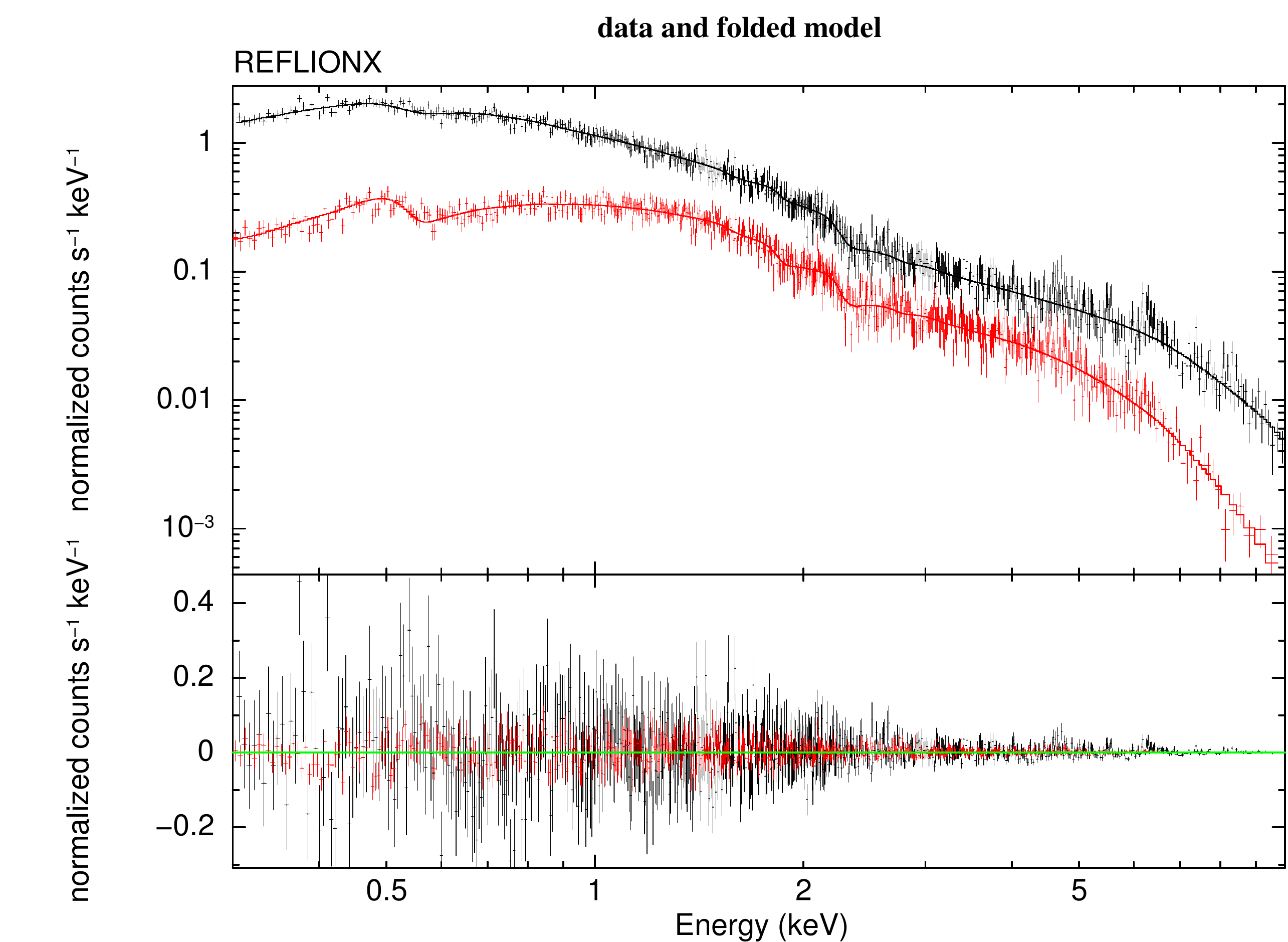} &
\includegraphics[width=.36\textwidth, angle=-90]{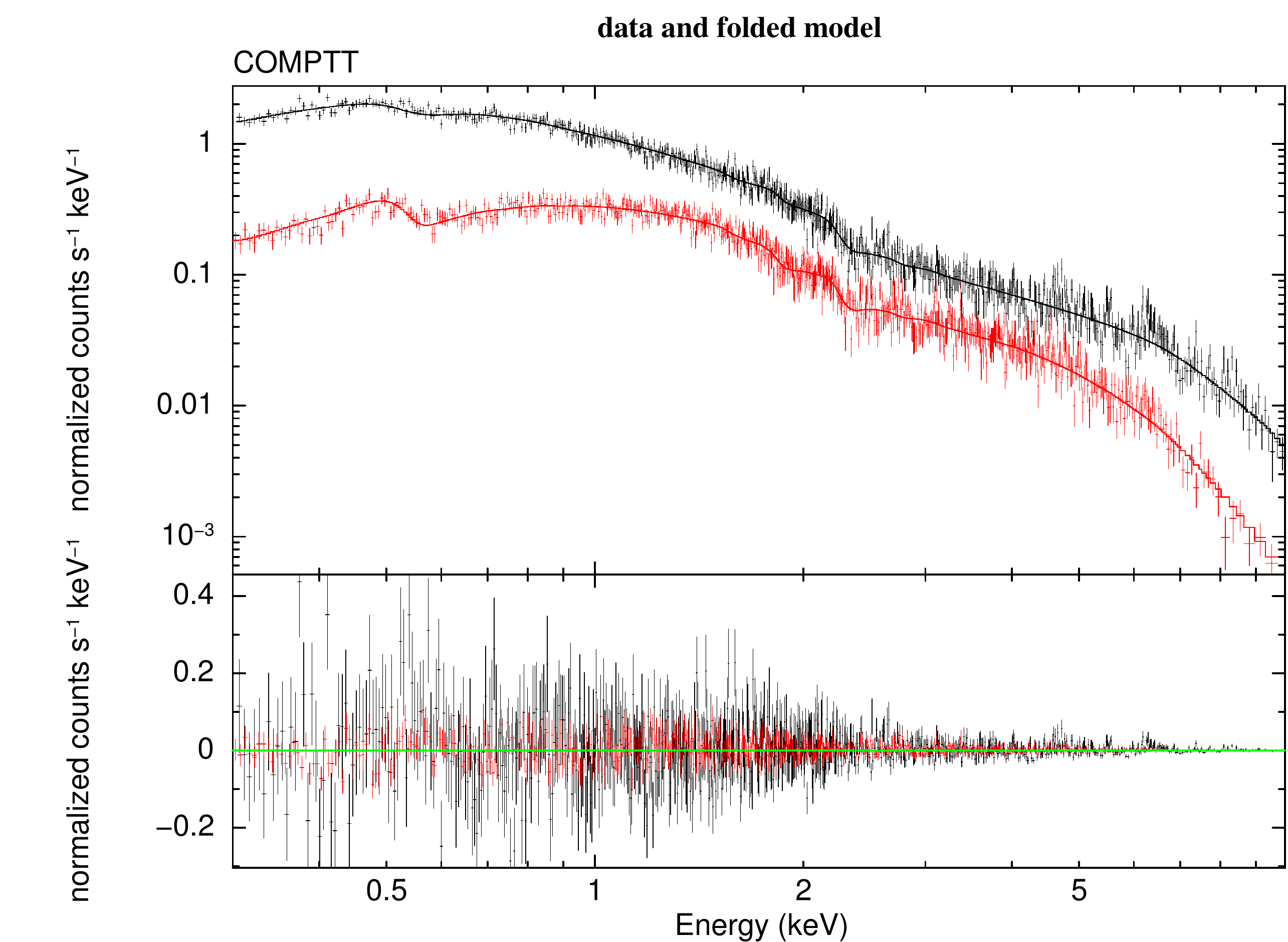} \\
\end{tabular}
\caption{Upper panels: The combined EPIC PN (black upper) and MOS (red lower) spectra of Zw 229.015 fitted with a power-law $and$ diskbb model (left); power-law $and$ swind1 model (right).
Lower panels: The combined EPIC PN (black upper) and MOS (red lower) spectra of Zw 229.015 
fitted with a power-law $and$ relativistically blurred reflection model (left); power-law $and$ comptonisation model (right)}
\label{fig2}
\end{figure*}

The parameters of the model are emissivity index of the disc, ionization parameter, inclination of the disc, iron abundance of the accreting matter (elements higher than iron are considered to have solar abundance), spectral index of the power-law radiation, inner and outer radii of the disc. For the inclination, we fix a modest value of $30\degree$ since Seyfert 1 galaxies generally have low inclination angles. The outer radius is fixed at $400 R_{\mathrm{g}}$ ($R_{\mathrm{g}}
={GM}/c^{2}$, where $G$ is the Newton's gravitational constant, $M$ is the black hole mass 
and $c$ is the speed of light) while the inner radius is kept free. All model parameters and their values are shown in Table \ref{table:Best-fit parameters}. The $\chi^2/dof$ comes out to be 1436/1404. Figure \ref{fig2} (lower panel, left) shows the fitted spectrum and the residual.

\subsection{Thermal comptonisation Model}\
This model was developed by \citet{1994ApJ...434..570T}, it is an analytic model that describes the comptonisation of soft photons in a hot plasma. The model assumes that seed photons emanating from the accretion disc are Compton-boosted by plasma in an optically thick medium i.e., the thermal disc supplies energy for the Compton tail. The model includes relativistic effects and the approximations used work well for both optically thin and thick regimes. In this model, the so-called $\beta$-parameter (which has no dependence on geometry) and the plasma temperature fully describe the comptonised spectrum. Subsequently the optical depth is determined for a given geometry as a function of $\beta$. Parameters of the model include the photon index of the power-law component, the temperature of the seed photons, the optical depth and temperature of the comptonising plasma. The complete list of parameters and their fit values are provided in Table \ref{table:Best-fit parameters}. The model is implemented as COMPTT in XSPEC. We fix the seed photons temperature to $30 \,\mathrm{eV}$ since this is about the typical temperature for the inner region of a Shakura-Sunyaev accretion disc around a black hole of mass $10^7M_{\odot}$. After fitting, the $\chi^2/dof$ value comes out to be 1435/1407. The model spectrum is shown in Figure \ref{fig2} (lower panel, right).

\section{Timing Analysis}\label{sec:sec4}
We extract lightcurves in the soft energy band $0.3-1.0\,\mathrm{keV}$ and hard bands $1.0-10.0\,\mathrm{keV}$ and $3.0-10.0\,\mathrm{keV}$. This is based on the shape of the spectrum with bin times of $100\,\mathrm{s}$ and $500\,\mathrm{s}$ in order to study the correlation between the soft and hard energy bands. Figure \ref{fig3} (upper panel) shows the lightcurves of the source in two of the mentioned energy bands as well as the lightcurve from the hardness ratio (HR) defined as the ratio of the flux in the $3.0-10.0\,\mathrm{keV}$ to that in the $0.3-1.0\,\mathrm{keV}$ in this case. The $0.3-1.0\,\mathrm{keV}$ band shows a trough-to-peak variation by a factor of $\sim2.4$, while the $3.0-10.0\,\mathrm{keV}$ band shows a trough-to-peak variation by a factor of $\sim3.6$, further corroborating the fact that the X-ray emissions from Zw 229.015 is highly variable. To investigate the possibility of a lag between the lightcurves of the soft ($0.3-1.0\,\mathrm{keV}$) and the hard ($1.0-10.0\,\mathrm{keV}$ $and$ $3.0-10.0\,\mathrm{keV}$) energy bands, we carry out cross-correlation analysis on the lightcurves. Afterwards, to search for signatures of low dimensional chaos (important for studying the disc dynamics) in the optical as well as X-ray variability of the source, we carry out nonlinear timing analysis \citep{1983PhRvA..28.2591G} on the source.

\subsection{Cross-Correlation Analysis}
\begin{figure*}
\centering
\begin{tabular}{@{}cc@{}}
\multicolumn{2}{c}{\includegraphics[width=.35\textwidth, angle=-90]{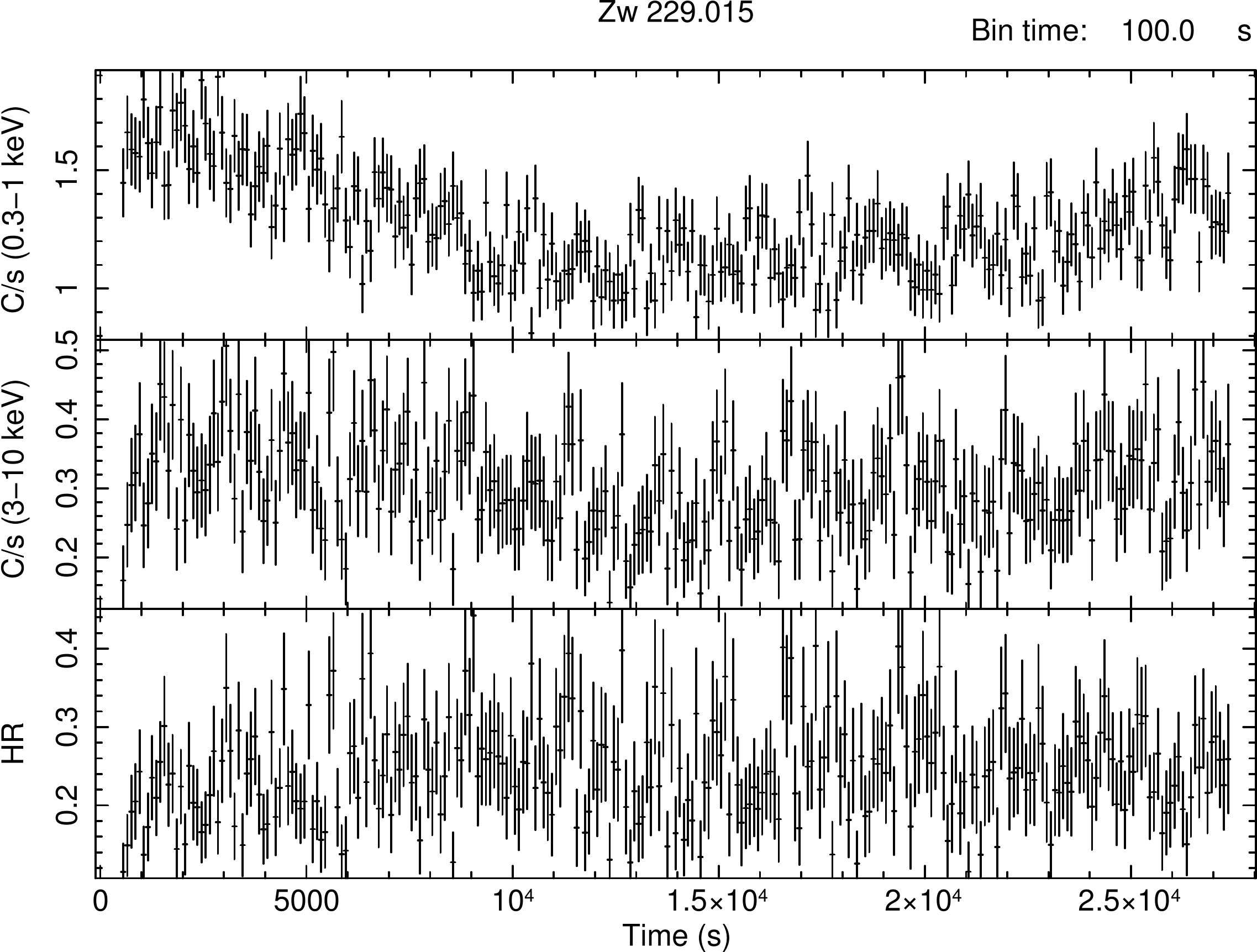}} 
\\
\includegraphics[width=.35\textwidth]{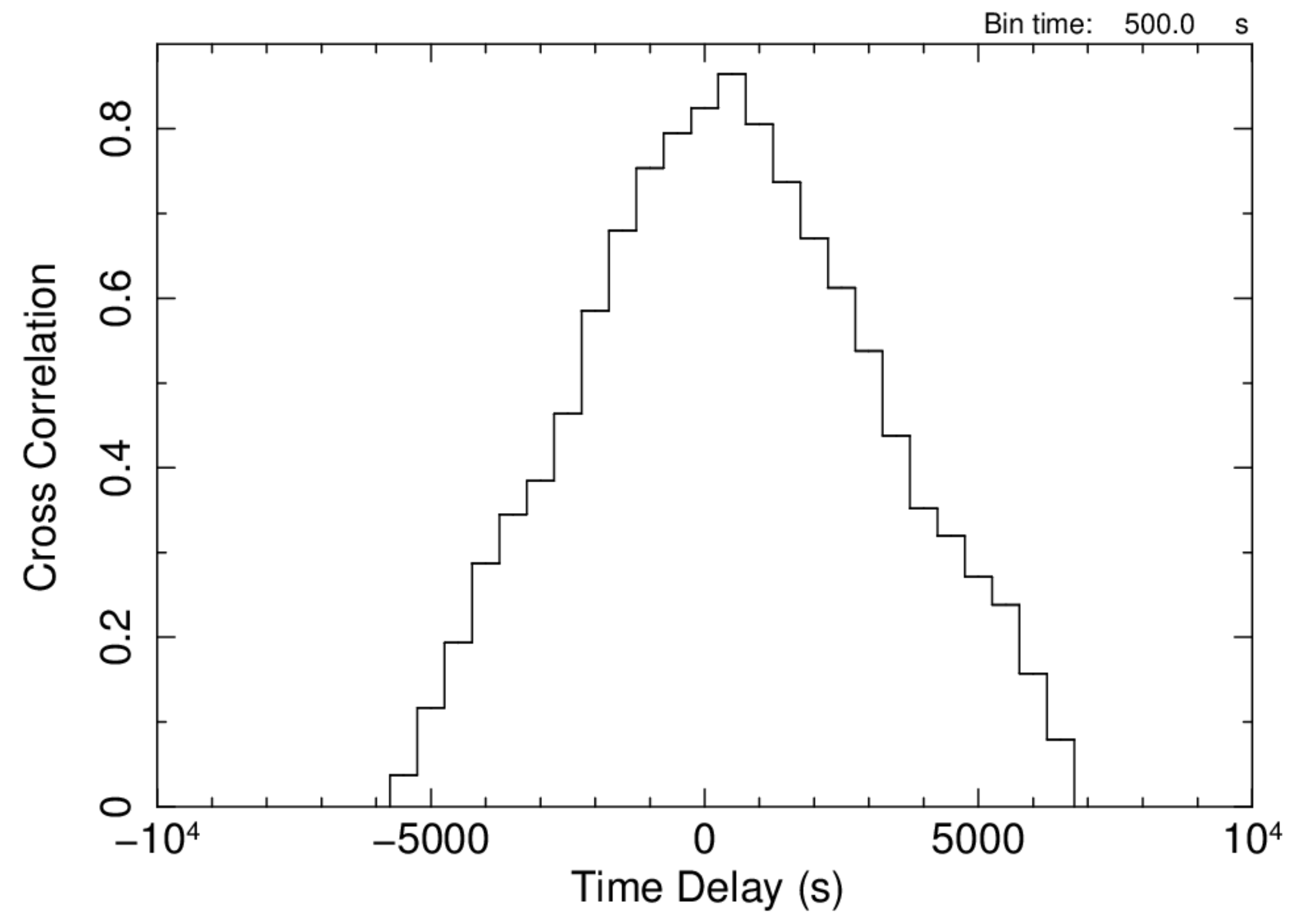} &
\includegraphics[width=.35\textwidth]{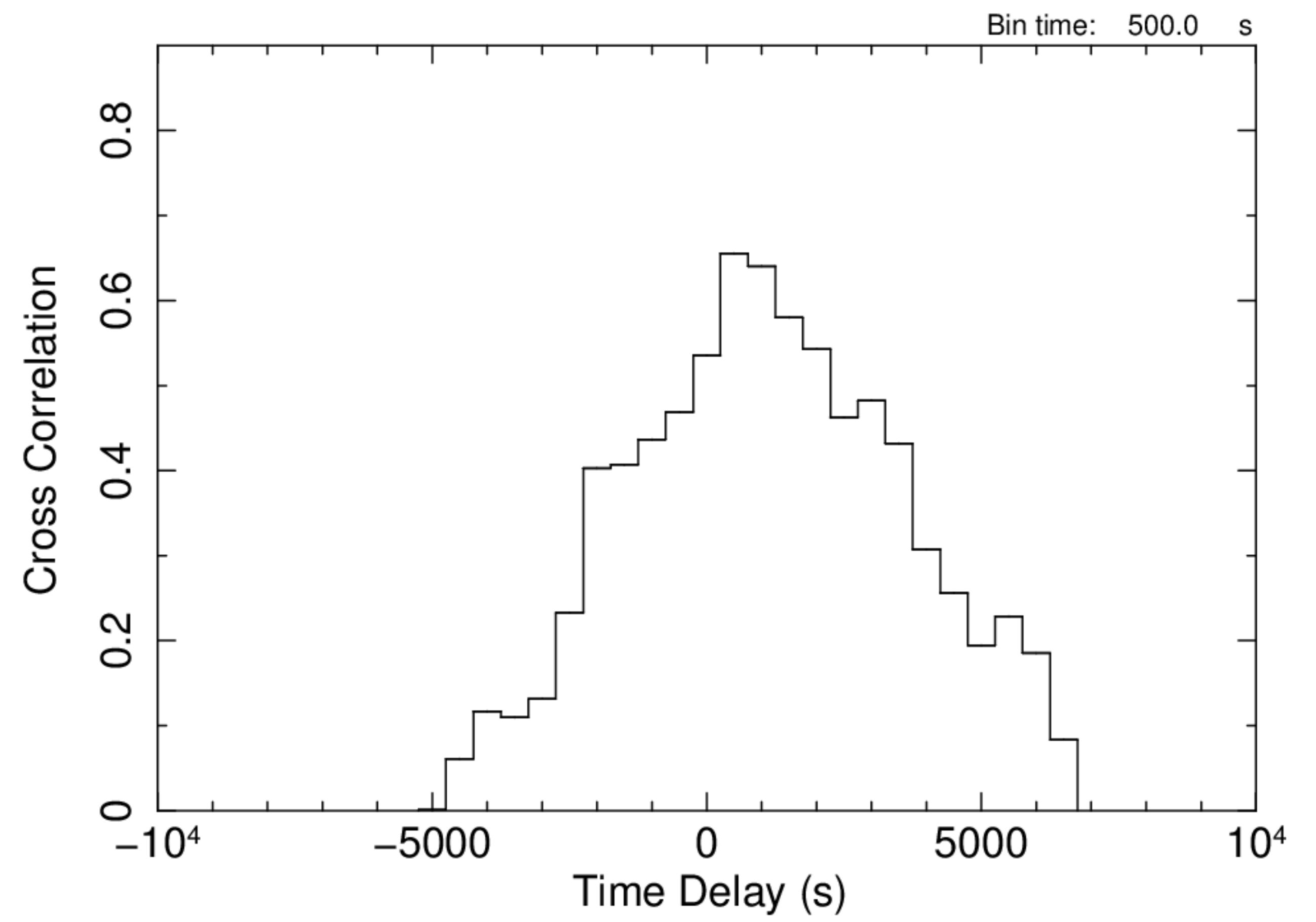} \\
\end{tabular}
\caption{Upper panel: Lightcurves of the energy bands $0.3-1.0\,\mathrm{keV}$ and $3.0-10.0\,\mathrm{keV}$ as well as the Hardness ratio.
Lower panels: CCF as a function of lag obtained using $\chi^2$-minimisation technique in XRONOS for the $0.3-1.0\,\mathrm{keV}$ $and$ $3.0-10.0\,\mathrm{keV}$ variability (left) and $0.3-1.0\,\mathrm{keV}$ $and$ $1.0-10.0\,\mathrm{keV}$ variability (right).
}
\label{fig3}
\end{figure*}

\
We use the XRONOS\footnote{\url{http://heasarc.gsfc.nasa.gov/docs/xanadu/xronos/xronos.html}} programme \textit{crosscor} to compute the cross-correlation function (CCF) between the soft and hard band lightcurves with $500\,\mathrm{s}$ bins. The variations of CCFs as 
functions of time delay between the soft and hard energy bands are shown in Figure \ref{fig3} 
(lower panels). In the CCF, visual inspection shows that the most prominent peaks in both 
the curves are at a positive time lag, implying that the hard band lags the soft \citep[see e.g.,][]{2015MNRAS.453.2877H}. A time lag between the variability of the soft and hard X-ray bands has been reported for some AGNs before now \citep[e.g.,][]{2004MNRAS.347..269G,2006ApJ...651L..13D,2016MNRAS.457..875P} implying plausibly different emission regions for the soft excess and the hard power-law for Seyfert 1  AGNs.

\begin{figure*}
\centering
\begin{tabular}{@{}cc@{}}
\includegraphics[width=.39\textwidth]{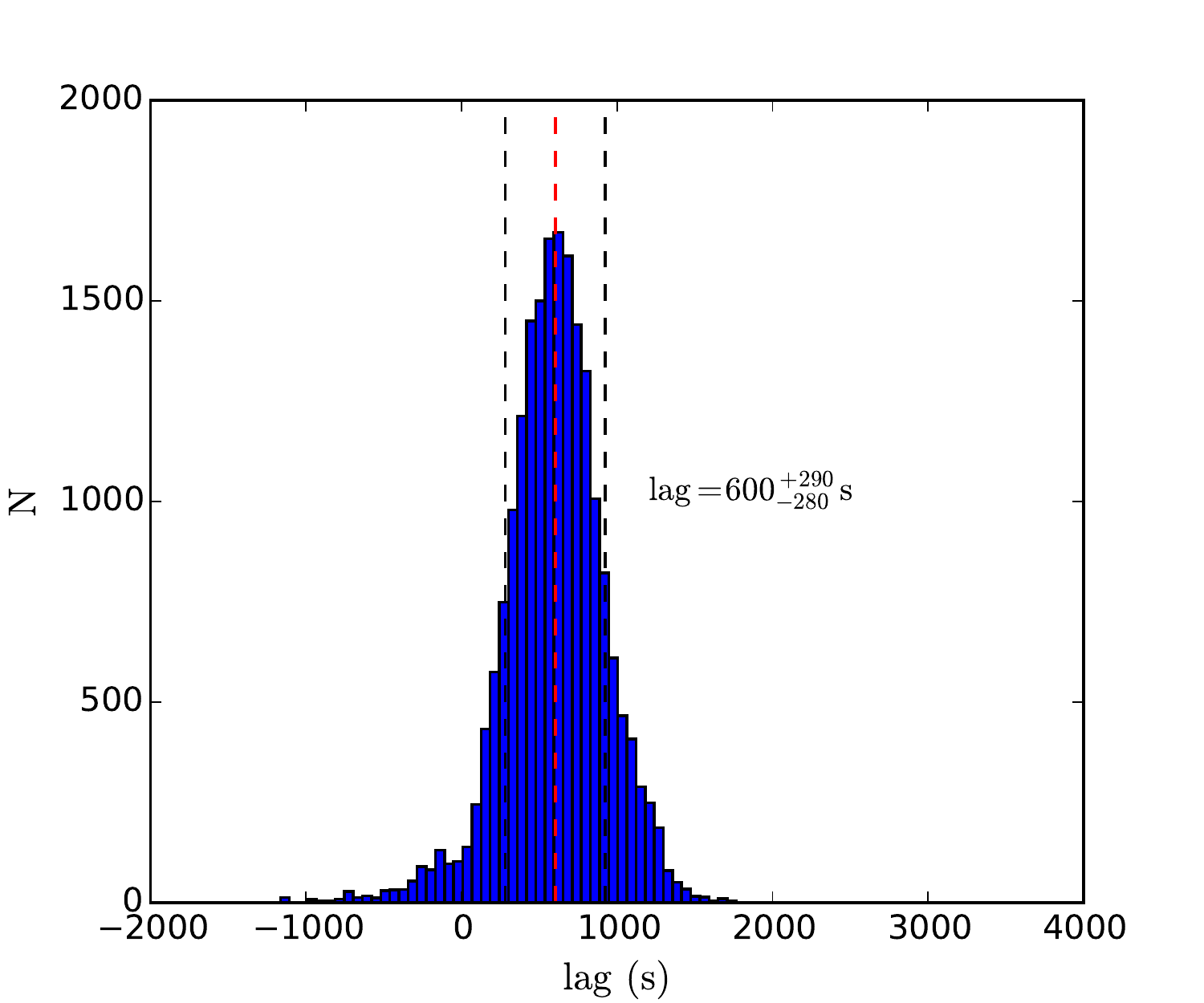} &
\includegraphics[width=.39\textwidth]{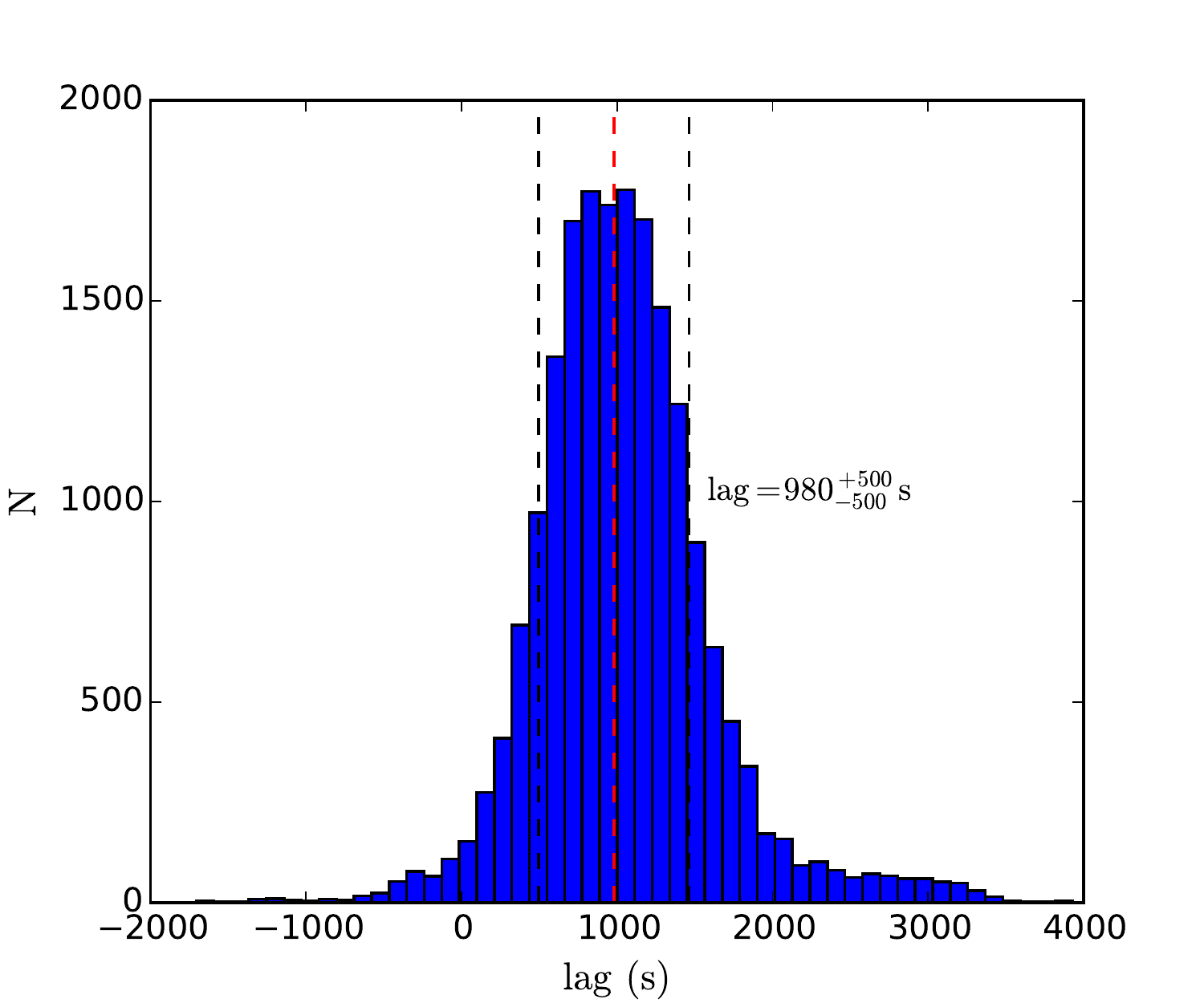} \\
\\
\includegraphics[width=7cm, height=6cm]{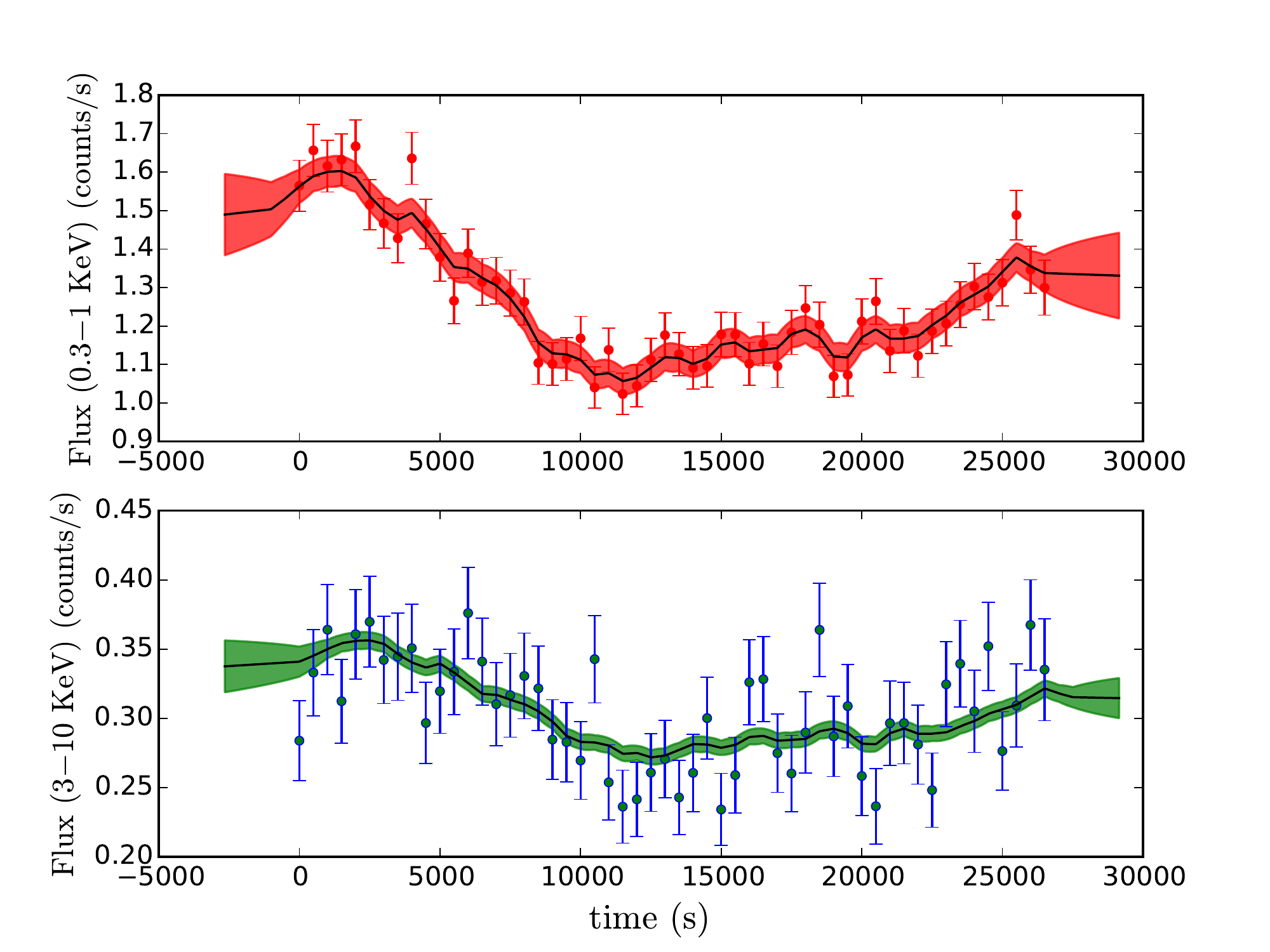} &
\includegraphics[width=7cm, height=6cm]{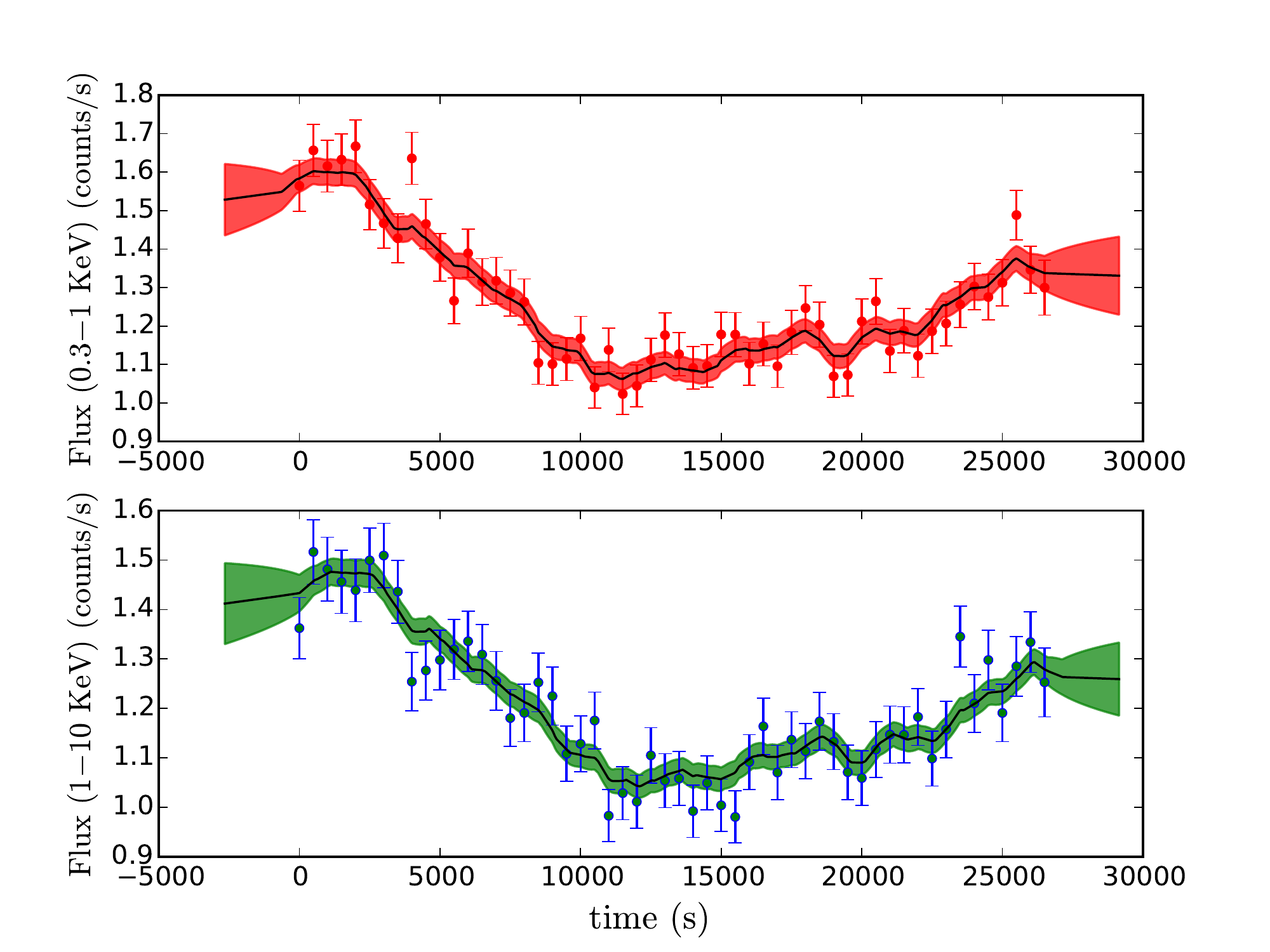} \\
\end{tabular}
\caption{Upper panels: Histogram showing the probability distribution of JAVELIN lag between the $0.3-1.0\,\mathrm{keV}$ $and$ $3.0-10.0\,\mathrm{keV}$ (left) and $0.3-1.0\,\mathrm{keV}$ $and$ $1.0-10.0\,\mathrm{keV}$ (right) X-ray variabilities.
Lower panels: Respective lightcurves of different bands obtained using the JAVELIN code.
}
\label{fig4}
\end{figure*}

As mentioned by \citet{2007ApJ...671.1284D}, due to the complex nature of the shape of the CCF, the non-Gaussian distribution of the errors and the inter-dependence of the CCF data points, it is somewhat difficult to measure the statistical significance of the time lag from model fitting to the CCF using the $\chi^{2}$ minimisation technique. As a result, to estimate the lag and its uncertainty between soft and hard bands, we use the JAVELIN\footnote{\url{http://www.astronomy.ohio-state.edu/~yingzu/codes.html}} code which is a Python implementation of the SPEAR \citep[Stochastic Process Estimation for AGN Reverberation;][]{2011ApJ...735...80Z}. JAVELIN has been successful in predicting reliably time lags in AGN optical lightcurves. It uses a damped random walk \citep[DRW;][]{2009ApJ...698..895K,2013ApJ...765..106Z} process to model the continuum emission in an AGN and then model the emission line lightcurve which is a smoothed, scaled and shifted version of the former. It estimates the best fit lag by comparing model lightcurves with the observed ones. After fitting this code to the data, we obtain a positive time lag of ${600^{+290}_{-280}}\,\mathrm{s}$ between the $0.3-1.0\,\mathrm{keV}$ and $3.0-10.0\,\mathrm{keV}$ flux variability and ${980^{+500}_{-500}}\,\mathrm{s}$ between the $0.3-1.0\,\mathrm{keV}$ and $1.0-10.0\,\mathrm{keV}$ flux variability, which implies that the hard X-ray lags the soft in its variability by an amount of this order. Significantly low signal to noise ratio (SNR) does not allow us to explore softer and harder bands. The plots of probability distribution functions of the lags as obtained from JAVELIN fit are shown by the histogram in Figure \ref{fig4} (upper panels), while the lower panels show the best fit lightcurves of different bands as obtained from the code. The relatively large uncertainties are possibly due to the data quality. Worthy of note is the fact that the results obtained through \textit{crosscor} and the  JAVELIN code are in good agreement. This leads us to the conclusion that the lag will not be more than $\sim1000\,\mathrm{s}$. This observed time lag definitely puts a further constraint on the tenable model to explain the origin of the soft excess.

\subsection{Nonlinear Timing Analysis}\
We apply the nonlinear time series analysis technique to search for signatures of low dimensional chaos in the X-ray and optical lightcurves of Zw 229.015 using its \textit{XMM-Newton} and \textit{Kepler} data respectively. We use the CI method.

One of the standard methods for reconstructing the dynamics of a nonlinear system from a time series is the delay embedding technique \citep{1983PhRvA..28.2591G}. The idea is to construct a CI equal to the probability that two arbitrary points in phase space are closer together than some separation $R$. Since the dimension of the system (i.e., the number of governing equations or variables) is not a priori known, one has to construct the dynamics for different embedding dimensions $M$. In this technique, vectors of dimension $M$ are created from the time series $s(t_{i})$ using a delay time $\tau$ such that
\\
\begin{equation} 
x(t_{i})=[s(t_{i}), s(t_{i}+\tau), s(t_{i}+2\tau),...,s(t_{i}+(M-1)\tau)].\\
\end{equation}
Typically, $\tau$ is suitably chosen such that the vectors are not correlated, i.e., when the autocorrelation function of $x(t_{i})$ approaches zero or when it attains its first minima. 

\textit{Correlation Dimension}: This provides a quantitative picture of the reconstructed dynamics. The computational procedure involves choosing a large number of points in the reconstructed dynamics as centres. The correlation integral $C_{M}(R)$ is the number of points that are within the distance $R$ from the centre averaged over all the centres and is written as
\begin{equation}
C_{M}(R)=\frac{1}{N({N_{c}-1})}\mathop{\sum^{N}\sum^{N_{c}}}_{i=1\ j\neq i\ j=1}{H(R-|{x_{i}-x_{j}}|)},
\end{equation}
where $x_{i}$, $x_{j}$ are the reconstructed vectors, $H$ is the Heaviside step function, $N$ is the number of points and $N_{c}$ the number of centres. The correlation dimension $D_{2}$ is essentially the scaling index of the variation of $C_{M}(R)$ with $R$ and it is expressed as 
\begin{equation}
D_{2}=\mathop {\lim }\limits_{R \to 0}(\frac{d\mathrm{log}C_{M}(R)}{d\mathrm{log}R}\big). \\
\end{equation}

Knowledge of $D_{2}$ allows one to determine the effective number of differential equations describing the dynamics of the system in principle. More details on this technique and its interpretation can be found in the works of \citet{2006ApJ...643.1114M} and \citet{2015A&A...576A..17B} to mention but a few. $D_{2}$ can be calculated from the linear part of the plot of log$[C_{M}(R)]$ against log$[R]$ and its value depends on the value of $M$. The plot of $D_{2}$ against $M$ can reveal the nonlinear dynamical properties of the system. If $D_{2} \approx M$ for all $M$, then the system is stochastic, however for a deterministic system, initially $D_{2}$ increases linearly with $M$ until it reaches a certain value and saturates. This saturated value of $D_{2}$ is taken to be the correlation/fractal dimension of the system. Worthy of mention is the fact that a saturated $D_{2}$ is a necessary but not a sufficient condition for chaos. The existence of colour noise in a stochastic system may lead to a saturated $D_{2}$ of low value as well. Therefore, it is customary to analyse the data by an alternative approach to distinguish it from a pure noisy time series. To do this, one applies the surrogate data analysis technique \citep{1992PhyD...58...77T}. Simply put, surrogate data are random data generated by taking the original signal and reprocessing it so that it has the same distribution and power spectrum as the original data but has lost all its deterministic characteristics (i.e., phase randomisation). Then the same analysis is carried out on the surrogates to identify any distinguishing features. The scheme proposed by \citet{1996PhRvL..77..635S} known as the Iterative Amplitude Adjusted Fourier Transform (IAAFT) is more consistent to generate surrogates.

\begin{figure}
\centering
\includegraphics[width=0.35\textwidth, height=0.20\textheight]{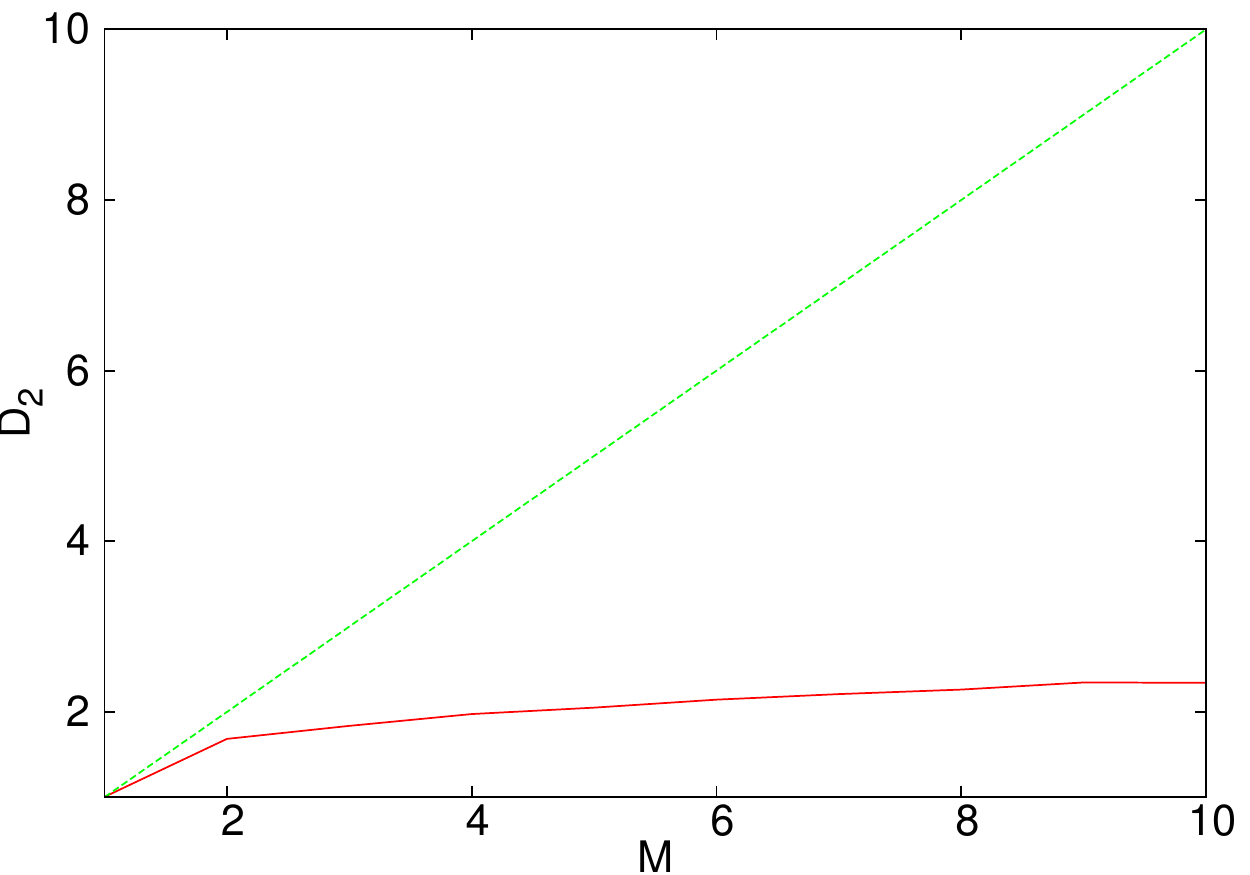}
\caption{The variation of correlation dimension as a function of embedding dimension for the Lorenz system showing saturation at low $D_{2}$ (red solid line). The green dashed line along the diagonal of the figure indicates an ideal stochastic curve.}
\label{fig5}
\end{figure}

\begin{figure}
\centering
\includegraphics[width=.35\textwidth, height=0.20\textheight]{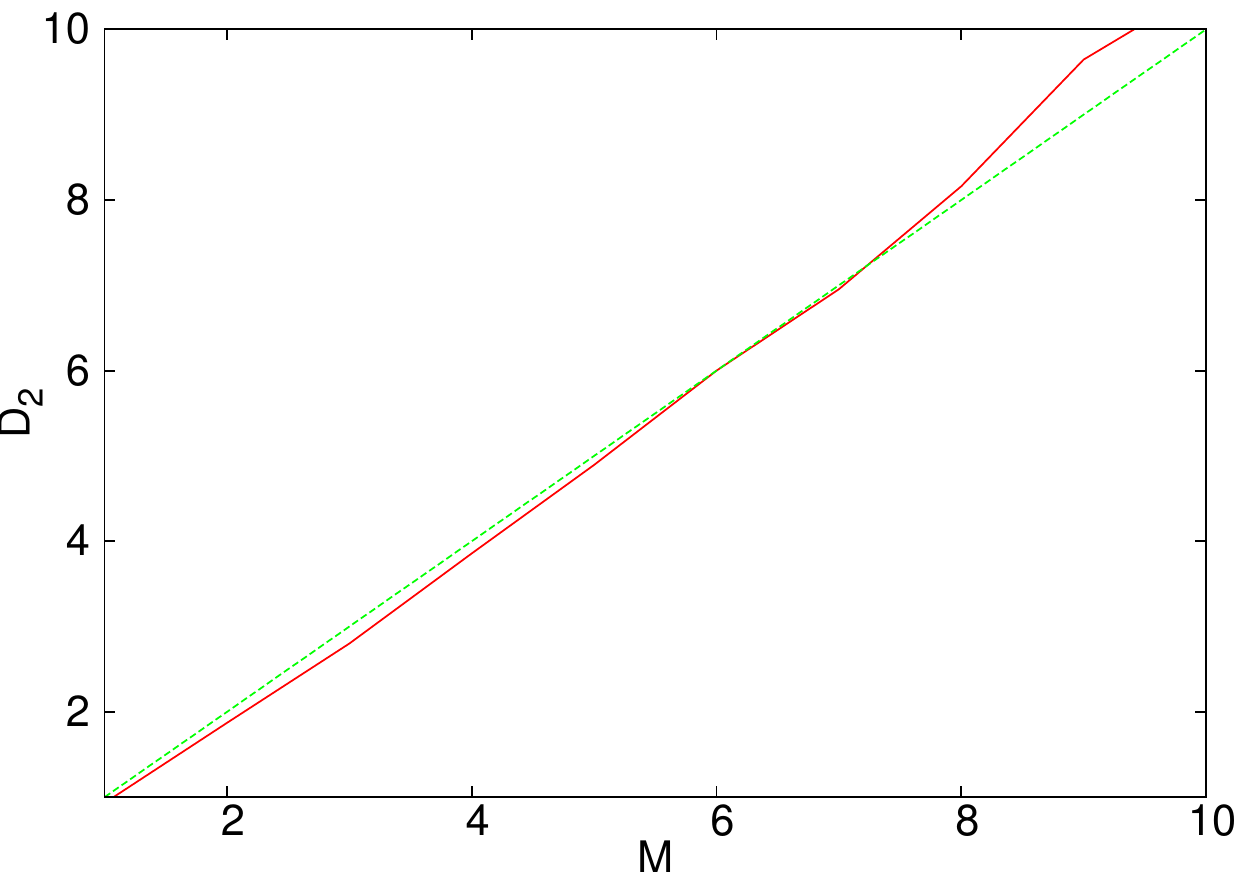}
\caption{The variation of correlation dimension as a function of embedding dimension from \textit{XMM-Newton} EPIC-PN data showing the stochastic nature of its X-ray lightcurve (red solid line). The green dashed line shows the behaviour of an ideal stochastic system.}
\label{fig6}
\end{figure}

\begin{figure}
\centering
\includegraphics[width=.35\textwidth, height=0.20\textheight]{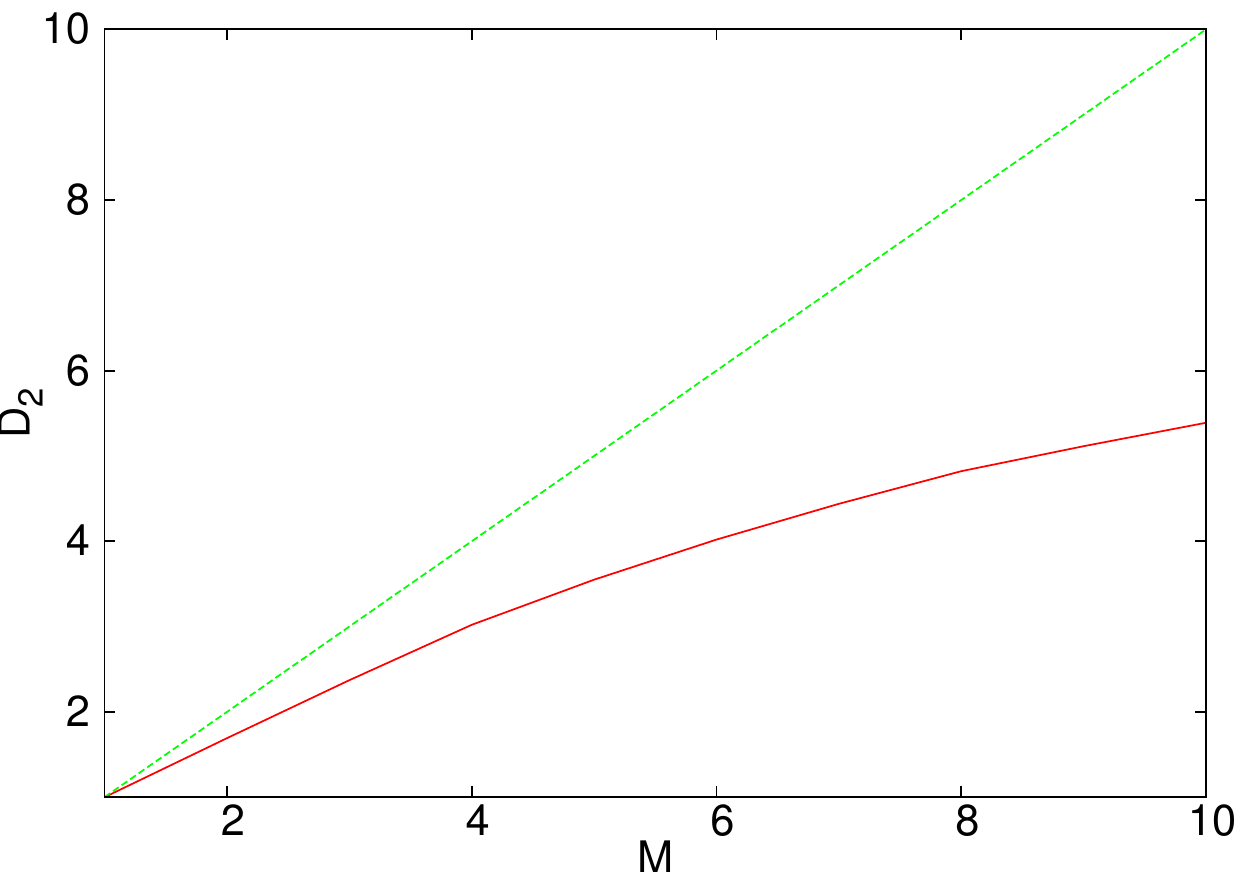}
\caption{ The variation of correlation dimension as a function of embedding dimension from the \textit{Kepler} short cadence data, it reveals the stochastic nature of the optical lightcurve of the system (red solid line). The green dashed line shows the behaviour of an ideal stochastic system.}
\label{fig7}
\end{figure}

Before applying the CI method to real data, we apply it to the lightcurve from the Lorenz system which is known to exhibit fractal dimension $(D_{2} = 2.06)$ for a particular parameter $\mathrm{R} = 28$ (not same as $R$ used in the $C_{M}(R)$ equation). Figure \ref{fig5} confirms that the Lorenz system is indeed chaotic, showing the variation of correlation dimension as a function of embedding dimension. On applying the CI method to the \textit{XMM-Newton} and the \textit{Kepler} data, no saturation is observed in the variation of $D_{2}$ with $M$ as shown in Figures \ref{fig6} and \ref{fig7} respectively. While the 
\textit{Kepler} data exhibit $D_2$ deviating from the ideal stochastic curve, $D_2$ 
for \textit{XMM-Newton} data practically overlap with that of stochasticity.
We do not consider including surrogate data here since the original lightcurves do not 
show signatures of chaos. More so, the $D_2-M$ curves for surrogates and the original data from the \textit{Kepler} lightcurve overlap (not shown in Figure \ref{fig7}).
Based on this analysis, we infer that the X-ray as well as optical lightcurves of Zw 229.015 are dominated plausibly by stochastic noise rather than deterministic chaos. The implication of this could be that advection is an important component of the accretion flow dynamics with random cooling processes \citep[general advective accretion flow;][]{2010MNRAS.402..961R} destroying any signature of self-similarity in the flux variability, making it stochastic.

\section{Discussions}\label{sec:sec5}
We have analysed the archival \textit{XMM-Newton} and \textit{Kepler} photometric data  for Zw 229.015 with the aim of understanding the spectral and timing behaviours of the source and to search for possible correlation between them. The X-ray energy range considered here is $0.3-10.0 \,\mathrm{keV}$ which is fairly well fitted with a simple power-law except for the presence of some soft excess noticed below $1.0 \,\mathrm{keV}$. 

\subsection{The soft excess}\
In order to understand the origin of the soft excess, motivated by the work of \citet{2007ApJ...671.1284D}, \citet{2014MNRAS.440..106B} and several others, we apply four models in fitting the EPIC-PN and MOS spectra of the source. The models are multicolour disc blackbody model, smeared wind absorption model, relativistically blurred reflection model and thermal comptonisation model. Worthy of mention is the fact that the smeared wind absorption model and the relativistically blurred reflection model relate the soft excess to atomic processes in the disc.

The disc temperature that we have obtained from the diskbb fit is $156 \,\mathrm{eV}$ which is much higher than the predicted value for an optically thick, geometrically thin accretion disc around a black hole of mass $\sim10^7M_{\odot}$. From standard accretion disc theory \citep{1973blho.conf..343N,1973A&A....24..337S}, the approximate temperature of the disc blackbody at a radius $r$ is given by
\begin{equation}
 T_{\mathrm{eff}}(r) \sim 6.3 \times 10^{5} \Big(\frac{\dot{M}}{{\dot{M}}_{\mathrm{Edd}}}\Big)^{1/4} \Big(\frac {M}{10^{8} M_{\odot}}\Big)^{-1/4} \Big(\frac{r}{r_{\mathrm{s}}}\Big)^{-3/4} ~{\rm K},
\end{equation}
where $\dot{M}$ is the black hole accretion rate, ${\dot{M}}_{\mathrm{Edd}}$ is the Eddington accretion rate, $\dot{M}/{\dot{M}}_{\mathrm{Edd}}$ is the scaled accretion rate and $r_{\mathrm{s}}$ is the Schwarzschild radius of the black hole given as $r_{\mathrm{s}}=2R_{\mathrm{g}}$. Thus for a non-spinning black hole of this mass, taking $r = 3r_{\mathrm{s}}$, the value of the disc temperature will be $\sim 30 \,\mathrm{eV}$. \citet{2012MNRAS.420.1848D} proposed the idea of colour temperature correction to AGN disc spectrum but eventually they came up with the conclusion that even though the colour temperature correction is substantial, it is not sufficient to account for the origin of the soft excess. Furthermore, the temperature of this component is fairly constant and not related to the luminosity and mass of the black hole \citep{2004MNRAS.349L...7G}. However, the disc model predicts that the disc temperature should be correlated with the black hole mass. Thus, the possibility of Compton scattering of photons could be important. Within the observed energy band, the spectrum does not show any cut-off which implies that it is not possible to constrain the temperature of the corona, as such observing the source in hard X-rays may help to ascertain the thermal Compton origin of the power-law.

The Smeared wind absorption model of \citet{2004MNRAS.349L...7G} posits that the soft excess emission is predominantly a consequence of strong and smeared absorption near $\sim0.7\,\mathrm{keV}$. The idea is that the soft excess and the hard power-law are from the same continuum component but for the small contributions from smeared emission lines. This means, the soft and hard band emissions are from the same continuum component or physical process - Comptonisation \citep{2007ApJ...671.1284D}. However, there is a major problem with this absorption model which is that the predicted smearing velocity for the source ($\sim0.4c$) is unphysically large for disc winds. As shown by hydrodynamic simulations, the line driven winds do not have the required smearing velocity \citep[e.g.,][]{2009ApJ...694....1S}. It remains to be confirmed if magnetically driven winds can provide the required smearing \citep{2007ApJ...671.1284D}. Outflowing winds having velocities as high as $0.1-0.2c$ have been observed in a few sources with high accretion rates \citep[e.g.,][]{2003MNRAS.345..705P}, however, when we fix the smearing velocity to $0.1c$, we get unacceptably poor fits.  

In the reflionx model, we have considered the soft excess as having an atomic origin in which case a hot corona produces the power-law continuum by the comptonisation of soft photons from the disc. These hard power-law photons irradiate the disc producing fluorescent secondary emissions which are subsequently blurred by the strong gravity in the vicinity of the black hole and the high velocities of the accreting material. In the reflection model, substantial fraction of the soft X-ray excess emission is due to the blurred reflection from the partially ionized material owing to lines and bremsstrahlung from the hot surface layers \citep{2005MNRAS.358..211R}. According to \citet{2006MNRAS.365.1067C}, this model is best explained in the scenario of a small power-law illuminating region above the central black hole where light bending effects deflect the illuminating power-law radiation on to the disc and thus is not fully detected. Although there is no evidence for strong Fe K$_{\alpha}$ line in this source, such a line may be hidden in the strong X-ray continuum, which involves the reprocessing of the primary X-ray emission from the corona. This model naturally explains the constancy of the observed temperature seen in many sources which is a measure of atomic transition in this case as mentioned by \citet{2008A&A...482..499D}.

In the thermal comptonisation model, we have modeled the soft excess in the energy range $0.3-1.0 \,\mathrm{keV}$ by considering disc photon comptonisation in an optically thick thermal corona. The seed photon temperature is fixed at $30\, \mathrm{eV}$  (typical for a black hole mass $\sim10^7M_{\odot}$). Corona temperature of $\sim632 \,\mathrm{eV}$ and $\sim635 \,\mathrm{eV}$ are obtained for disc and spherical geometries respectively while the optical depth comes out to be 8.7 and 18.1 for disc and spherical geometries respectively. As noted by \citet{2016xsce.book.....N}, the soft X-ray excess can be considered to be an additional primary emission rather than a contamination from a host galaxy or a secondary emission component. From the values of optical depths and plasma temperature that we obtain from this model, we conclude that the soft excess is likely to be produced in a relatively cool and dense corona which is clearly different from those producing the dominant power-law in which case the plasma temperature is of the order of $\sim150 \,\mathrm{keV}$ and optical depth $< 1.0$. The positive lag we have obtained in the CCF analysis is in tandem with the prediction of the comptonisation model which is that the hard X-ray variation should lag the soft. The time lag in this case represents the difference in the photon escape times between the soft and the hard bands. This implies that the hard band photons tend to have undergone more scattering events and so travel longer and escape the corona later than the soft photons. Although, as reported by \citet{2004MNRAS.349L...7G} and \citet{2006MNRAS.365.1067C}, one caveat of this model is the relative constancy of the disc temperature (around $0.1-0.2\,\mathrm{keV}$), regardless of the central object's luminosity and mass. More samples of AGNs have to be studied to shed more light on this and from our result, comptonisation from a cold corona is plausibly a major contributor to the soft excess emission.

\subsection{Timing Analysis}
We have shown that the variability in the hard X-ray band lags the variability in the soft on a time scale of $\sim1000\,\mathrm{s}$. This is interesting because on the one hand, it reveals that they are most likely emitted from different regions of the accretion disc, a result that is consistent with the two corona model which posits that the soft X-ray photons may be the seed for the observed inverse comptonised hard photons. Based on this lag, we estimate the size of the corona system around the black hole as 
\begin{equation}
d=c\times t_{\mathrm{lag}},
\end{equation} 
where $d$ is the separation between the two coronae (i.e., the approximate extent of the coronae geometry) and $t_{\mathrm{lag}}$ is the measured lag. This gives a value of $d\sim3\times 10^{13}\mathrm{cm}=20R_{\mathrm{g}}$. This indicates that the coronae system is very compact. Thus it is not surprising that the X-ray luminosity of the source is only $\sim10^{43}\mathrm{erg\, s^{-1}}$ while beyond $20R_{\mathrm{g}}$, the disc is dominated by optical/UV emission. Although lags in the opposite sense have been reported on shorter time scales in some AGNs \citep[e.g.,][]{2011MNRAS.418.2642Z,2011MNRAS.416L..94E,2011ApJ...736L..37T,2011MNRAS.417L..98D,2012MNRAS.419..116F,2013MNRAS.431.2441D}, this is not evident in our analysis.

We have shown that Zw 229.015 exhibits stochastic variability as seen from both its X-ray and optical lightcurves. The stochastic nature of the X-ray variability may be explained possibly by invoking the light bending effects. This has a strong influence on the brightness of the primary power-law source (and by extension, the reflected component) making it appear faint to a distant observer when it is close to the black hole and bright when further away. In this case we consider the variability of the emitted X-ray photons as plausibly intrinsic to the corona or due to the position of the corona relative to the hole (which may be a stochastic rather than a deterministic effect). The optical emission on the other hand is known to be produced from the optically thick, geometrically thin accretion disc and according to \citet{2015A&A...576A..17B}, one way to account for the apparently stochastic optical lightcurve is to invoke a large number of active zones or emitting blobs. This implies that the overall lightcurve will be stochastic in nature irrespective of the exact injection mechanism in each zone. This is based on the assumption that deterministic injection will lead to deterministic optical variability, evidence for which we do not find. Moreover, the presence of significant advection component with random cooling within the flow is possible and this is to an extent corroborated by the value of $L_{\mathrm{bol}}/L_{\mathrm{Edd}}$ obtained for this source. The presence of such a component may result in random flux fluctuations and thus revealing the stochasticity seen in the X-ray lightcurve. It suffices to add that it is not surprising that both the X-ray and the optical variabilities of the source are stochastic since there is growing evidence that the X-ray and optical variations of Seyfert galaxies are correlated in some way \citep{2008MNRAS.389.1479A,2009MNRAS.394..427B,2016arXiv161201369P}.

\section{Summary}\label{sec:sec6}
Based on the result of our spectral and timing analyses of Zw 229.015, we have come up with the following conclusions.

\begin{enumerate}
\item The spectral analysis of the \textit{XMM-Newton} X-ray data of the source in the energy range $0.3-10.0 \,\mathrm{keV}$ reveals the presence of some soft excess emission below $1.0 \,\mathrm{keV}$ and a power-law dominance above $1.0 \,\mathrm{keV}$.

\item Of the four models with which we have fitted the spectra namely, multicolour disc blackbody, smeared wind absorption, relativistically blurred reflection and the thermal comptonisation models, the thermal comptonisation and relativistically blurred reflection models give the most acceptable spectral fits and moreover they offer more physical explanation to the origin of the soft excess emission.

\item From the observation of a positive time lag of $\sim1000\,\mathrm{s}$ between the soft and hard bands, we have inferred that the two are produced from different regions extending only up to $\sim20R_{\mathrm{g}}$ which gives credence to the possibility of a thermal componisation. We believe that future high sensitivity broad band measurements will hopefully give better constraints for Seyfert 1 AGNs.

\item From the nonlinear time series analysis of the source, both in the X-ray and optical bands, we do not find signatures of low dimensional chaos, thus buttressing the fact that X-ray and optical variabilities of Seyfert galaxies may be correlated in some way. Worthy of mention is the fact that although the CI method can provide an indication of nonlinearity, it may also fail to detect nonlinear behaviour. The test for the signature
of nonlinearity should be done by other methods as well in order
to have a more conclusive statement. For example, the nonlinear prediction 
error \citep{1997PhRvE..55.5443S} could be explored towards this mission, which we might indeed consider for future work. More so, because of the stochastic nature of the lightcurve variability of the source, we conclude that the source may exhibit a non-standard accretion flow (having plausibly a significant advection component) as evident also from its low bolometric luminosity \citep[$\sim0.05L_{\mathrm{Edd}}$;][]{2011ApJ...732..121B}. This is already observed in many black hole systems. These features may however change with time as the source is known to be rapidly variable. 
\end{enumerate}

\section*{Acknowledgments} We acknowledge the anonymous referee for comments that improved this manuscript. We wish to thank Ramesh Narayan, Gulab Dewangan and C. S. Stalin for their valuable comments over the manuscript and suggestions. A partial financial support from the project with research Grant No. ISTC/PPH/BMP/0362 is also acknowledged. 




\bibliographystyle{mnras}
\bibliography{ref}


\end{document}